\documentclass[11pt, letterpaper, logo, onecolumn, copyright]{main}

\usepackage[authoryear,sort&compress,round]{natbib}
\usepackage[most,breakable,skins]{tcolorbox}
\usepackage{microtype}
\usepackage{multirow}
\usepackage{xspace}

\definecolor{gblue9}{RGB}{23,78,166}
\definecolor{myCite}{HTML}{1C4587}
\hypersetup{
    citecolor=myCite,
    linkcolor=myCite,
    urlcolor=myCite
}

\makeatletter
\renewcommand{\verbatim@font}{\normalfont\fontsize{8.0}{7.6}\selectfont}
\makeatother
\newsavebox{\offlinepromptbox}
\newsavebox{\onlinepromptbox}
\newsavebox{\judgepromptbox}

\graphicspath{{figures/}}

\newcommand{\method}{MuseCritic\xspace}
\newcommand{\musedirect}{Muse-GRPO (Direct RM)\xspace}
\newcommand{\musegrpo}{Muse-GRPO (MuseCritic)\xspace}

\title{MuseCritic: Learning Multi-Aspect Song Rewards through Natural-Language Aesthetic Critiques}

\author{
  \textbf{Jiabao Zhuang}\textsuperscript{*,\dag},
  \textbf{Changhao Jiang}\textsuperscript{*},
  \textbf{Hanchen Wang}\textsuperscript{*},\\
  \textbf{Jiahao Chen}\textsuperscript{*},
  \textbf{Zhixiong Yang}\textsuperscript{*},
  \textbf{Zhenghao Xiang}\textsuperscript{*},\\
  Yifei Cao,
  Jiajun Sun,
  Hui Li,
  Ming Zhang,\\
  Tao Ji,
  Tao Gui\textsuperscript{\dag},
  Qi Zhang,
  Xuanjing Huang\\
  \vspace{0.3cm}
  \normalsize
  Fudan NLP Group, Fudan University\\
  \texttt{\normalsize 25213050071@m.fudan.edu.cn, tgui@fudan.edu.cn}
}

\begin{abstract}
Long-form song generation models continue to improve in duration, structural coherence, and acoustic complexity, increasing the need for reliable aesthetic rewards aligned with human preferences. However, reward models for complete songs remain limited, and existing evaluators typically predict scores in a single forward pass without readable explanations. To this end, we introduce MuseCritic, a semi-scalar reward model that generates a natural-language critique covering five aesthetic dimensions and uses it as an intermediate representation to predict continuous reward scores. MuseCritic follows a two-stage training pipeline: a teacher model first provides high-quality critiques for supervised fine-tuning, then the fine-tuned model generates its own critiques for reward learning, mitigating training-inference distribution shift. On an in-domain test set of 200 SongEval songs, MuseCritic reduces macro-averaged mean squared error from 0.2875 to 0.2316 and improves macro-averaged LCC, SRCC, and Kendall's $\tau$ to 0.9068, 0.8838, and 0.7178, respectively. On the out-of-domain Music Arena benchmark with 733 preference pairs, it achieves 71.35\% accuracy and remains competitive with strong music-specific reward models. Using MuseCritic with GRPO also improves Muse-0.6B on all nine aesthetic metrics from SongEval and Audiobox Aesthetics. These results show that critique-conditioned reward modeling reduces scoring error and provides an effective optimization signal for song generation. The project repository is available at \href{https://github.com/WuqnEl/MuseCritic}{\texttt{https://github.com/WuqnEl/MuseCritic}}.
\end{abstract}

\begin{document}

\begingroup
  \renewcommand\thefootnote{}
  \footnote{\hspace{-1.8em}\textsuperscript{*}Equal contribution.\quad
            \textsuperscript{\dag}Corresponding authors}
\endgroup

\maketitle

\section{Introduction}
\label{sec:introduction}

Text-to-music generation has progressed from synthesizing short clips to composing complete songs. Early systems generate musical excerpts from natural-language descriptions \citep{agostinelli2023musiclm,copet2023musicgen}, while recent systems---including DiffRhythm, YuE, LeVo, ACE-Step, and Muse---support minutes-long vocal music, lyric alignment, and fine-grained style control \citep{ning2025diffrhythm,yuan2025yue,lei2025levo,gong2025acestep,jiang2026muse}. As generation length and acoustic realism improve, reliable aesthetic evaluation becomes increasingly important for model selection and for providing optimization signals aligned with human preferences.

Existing evaluators, however, remain limited for complete songs. Objective metrics such as phoneme error rate, MuLan, and MuQ-MuLan capture lyric intelligibility or audio--text correspondence rather than perceptual and artistic quality \citep{huang2022mulan,zhu2025muq}. PAM, MusicEval, and Audiobox Aesthetics broaden evaluation to audio quality and musical aesthetics \citep{deshmukh2024pam,liu2025musiceval,tjandra2025audiobox}, while SongEval provides five expert aesthetic criteria for full-length synthetic songs \citep{yao2025songeval}; nevertheless, these evaluators map audio directly to scores and do not explain the audible evidence behind their judgments. In contrast, LLM-as-a-Judge and critique-conditioned reward models show that language models can interpret rubrics and generate corresponding rationales or supporting reasoning before scoring textual responses, thereby improving textual preference modeling \citep{zheng2023llmjudge,ankner2024cloud,yu2025criticrm,zhang2024genrm,liu2025deepseekgrm}. Whether this linguistic intermediate representation can improve multidimensional aesthetic reward modeling for complete songs remains underexplored. Existing audio reward models predict scores directly, while general-purpose multimodal models lack task-specific training and show weaker agreement with expert judgments.

\begin{figure*}[t]
    \centering
    \includegraphics[width=1.0\textwidth]{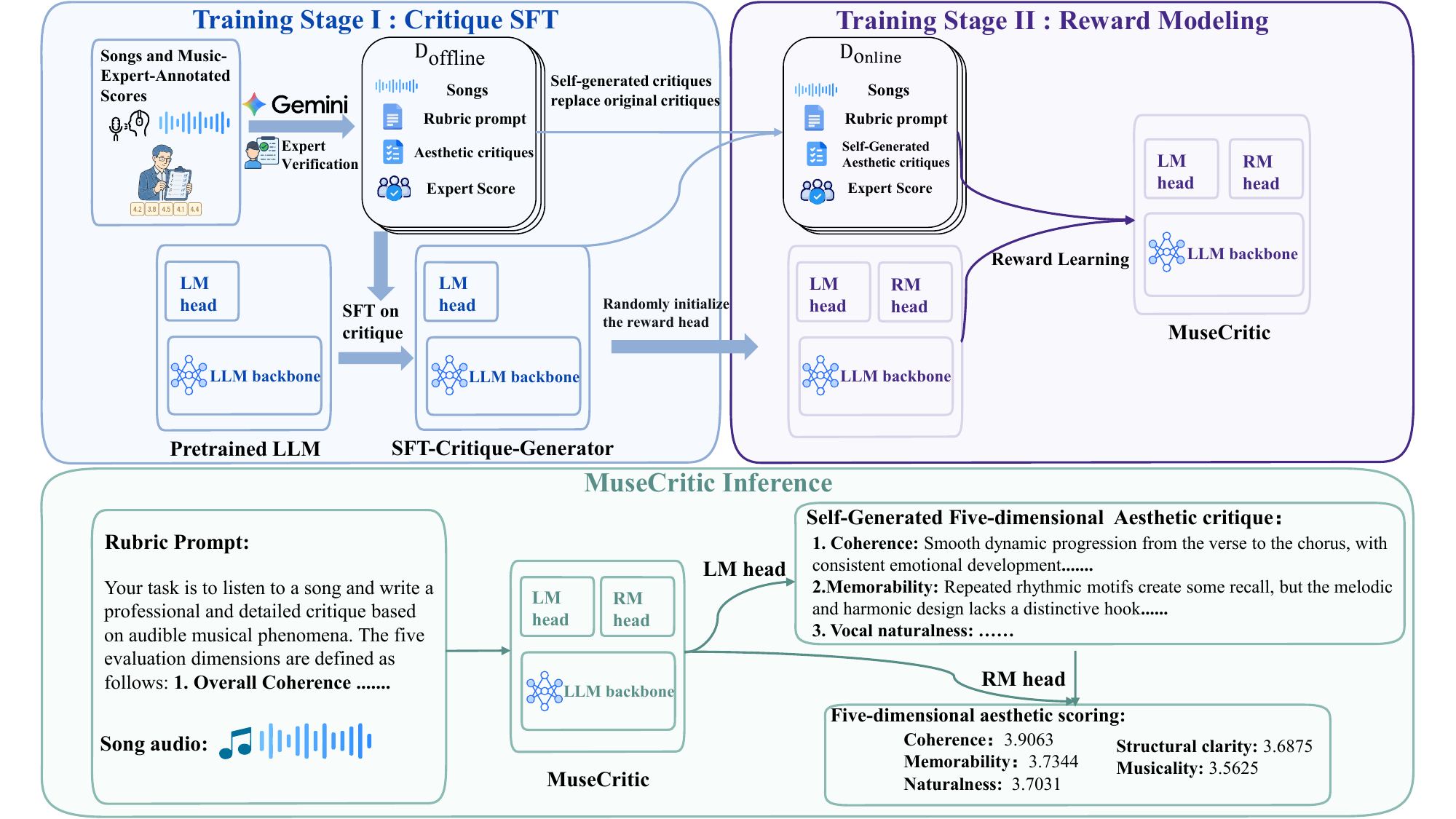}
    \caption{\textbf{Overview of the two-stage training and inference pipeline of \method.} In Stage~I, Gemini-3-Pro generates aesthetic critiques from songs, evaluation rubrics, and expert mean ratings to construct $\mathcal{D}_{\mathrm{off}}$, which is used to fine-tune a pretrained large language model into an SFT critique generator. In Stage~II, the SFT model replaces the teacher critiques with one self-generated five-aspect critique per song to construct $\mathcal{D}_{\mathrm{on}}$; a scalar reward head is then initialized, and \method is trained using expert mean ratings as regression targets. At inference time, \method first generates a five-aspect critique and then predicts five continuous aesthetic scores conditioned on the song, rubric, and self-generated critique.}
    \label{fig:overview}
\end{figure*}

In this work, we investigate explicit linguistic reasoning in song reward modeling, enabling the model to form a representation of aesthetic evidence aligned with the evaluation rubric before predicting multidimensional scores. To this end, we propose \method, a semi-scalar reward model built around natural-language aesthetic critiques. Given a song and a five-dimensional evaluation rubric, \method generates a single critique containing five labeled analyses: Overall Coherence, Memorability, Naturalness of Vocal Breathing and Phrasing, Clarity of Song Structure, and Overall Musicality. Conditioned on the song, rubric, and self-generated critique, a reward head then predicts five continuous aesthetic scores ranging from 1 to 5. The critique is not a post-hoc text appended to predicted scores; it is an explicit intermediate variable connecting song representations to rewards. This design introduces a variable-length textual representation before prediction and organizes auditory evidence distributed across a long song into dimension-aligned descriptions. Figure~\ref{fig:overview} summarizes the training and inference pipeline.

Across both in-domain and out-of-domain evaluations, critique-conditioned reward modeling consistently improves agreement with human judgments. On the held-out SongEval test set, \method reduces macro-averaged MSE from 0.2875 to 0.2316 and raises macro-averaged LCC, SRCC, and Kendall's $\tau$ to 0.9068, 0.8838, and 0.7178. On Music Arena, \method attains 71.35\% pairwise accuracy across 733 preference pairs and remains competitive with strong music-specific reward models. When used as the reward model for GRPO, it further improves Muse-0.6B on all nine aesthetic metrics reported by SongEval and Audiobox Aesthetics.

Our contributions are threefold:
\begin{itemize}
    \item We introduce \method, a semi-scalar reward model for complete songs. It generates a natural-language critique containing five dimension-specific analyses and uses this critique to predict five continuous scores, combining verbal evaluation and reward estimation within a single model (Figure~\ref{fig:overview}).
    \item We systematically evaluate \method on in-domain SongEval ratings and out-of-domain Music Arena preferences. Self-generated critiques improve agreement with expert scores and mitigate the critique distribution shift between reward training and inference (Tables~\ref{tab:in-domain}, \ref{tab:arena}, and~\ref{tab:ablation}).
    \item We use \method as the reward model for GRPO training of Muse-0.6B. The resulting policy improves on all nine observed metrics from SongEval and Audiobox Aesthetics, showing that the learned reward provides a useful optimization signal (Table~\ref{tab:downstream}).
\end{itemize}

\section{Related Work}
\label{sec:related}

\subsection{Text-to-Song Generation}

Language-conditioned music generation has progressed from short excerpts produced by systems such as MusicLM and MusicGen to complete songs with vocals and long-range structure \citep{agostinelli2023musiclm,copet2023musicgen}.
Recent systems use latent diffusion, dual-track tokenization, and semantic conditioning to generate minutes-long songs, while supporting lyric alignment, vocal--accompaniment coordination, and fine-grained style control \citep{ning2025diffrhythm,yuan2025yue,gong2025acestep,lei2025levo,jiang2026muse}.
MusicRL learns reward signals from human feedback, and LeVo applies multi-preference DPO to improve song quality and instruction following \citep{cideron2024musicrl,lei2025levo}.
These advances make reliable aesthetic evaluation and reward modeling increasingly important for model comparison and reinforcement learning.
Our work addresses this need by learning a reward model that both describes five-dimensional song aesthetics and predicts continuous scores.

\subsection{Reward Modeling and Music Aesthetic Evaluation}

Classical RLHF reward models fit human preferences with a scalar head on a pretrained language model, whereas LLM-as-a-Judge methods can generate evaluations and supporting rationales \citep{ouyang2022instructgpt,zheng2023llmjudge}.
Semi-scalar and generative reward models further incorporate critiques or reasoning traces before reward prediction \citep{ankner2024cloud,yu2025criticrm,zhang2024genrm,liu2025deepseekgrm}.
These methods primarily evaluate text responses; \method extends critique-conditioned reward modeling to long-form song audio and multidimensional aesthetic scores.

Music evaluators commonly predict proxy metrics or learned quality scores.
MuLan and MuQ-MuLan measure text--music correspondence \citep{huang2022mulan,zhu2025muq}; PAM, MusicEval, and Audiobox Aesthetics assess audio quality or musical aesthetics \citep{deshmukh2024pam,liu2025musiceval,tjandra2025audiobox}; and SongEval predicts five aesthetic dimensions for complete songs \citep{yao2025songeval}.
More recent models include the seven-dimensional SongBench evaluator, the scalar preference model TuneJury, and CMI-RM for musicality and instruction alignment \citep{wu2026songbench,kim2026tunejury,ma2026cmirewardbench}.
These evaluators output scalar or multidimensional scores without using a generated natural-language critique as an intermediate variable.
In contrast, \method generates rubric-aligned critiques and conditions its continuous reward predictions on the resulting textual evidence.

\section{Method}
\label{sec:method}

\subsection{Preliminaries}

\paragraph{Problem formulation.}
Let $x\in\mathcal{X}$ denote a complete song with vocals, $p$ an aesthetic-evaluation prompt that defines the assessment dimensions, and $K=5$ the number of dimensions.
The expert annotation is $\mathbf{y}^{*}=[y^{*}_{1},\ldots,y^{*}_{K}]\in[1,5]^K$.
The five dimensions are Overall Coherence, Memorability, Naturalness of Vocal Breathing and Phrasing, Clarity of Song Structure, and Overall Musicality \citep{yao2025songeval}.
We use an instruction-tuned audio language model $f_{\theta}$ as the shared backbone of our semi-scalar reward model.
Its critique generator $g_c$ and aesthetic scorer $g_s$ are
\begin{equation}
g_{c}=h_{\mathrm{LM}}\circ f_{\theta},\qquad
g_{s}=h_{\mathrm{RM}}\circ f_{\theta},
\end{equation}
where $h_{\mathrm{LM}}$ is a language-modeling head that generates a natural-language aesthetic critique, and $h_{\mathrm{RM}}$ is a reward-modeling head that produces $K$ continuous aesthetic scores. The two components always share the same backbone parameters $\theta$.

\paragraph{Reward modeling.}
Unlike reward models that predict a scalar directly from audio, the aesthetic scorer of \method jointly conditions on the song $x$, evaluation rubric $p$, and aesthetic critique $c$:
\begin{equation}
\hat{\mathbf y}=g_s(x,p,c).
\label{eq:reward-modeling}
\end{equation}
For any set $\mathcal S$ of expert-annotated samples, we learn the five-dimensional scores by minimizing mean squared error:
\begin{equation}
\begin{aligned}
\mathcal L_{\mathrm{MSE}}(\mathcal S;g_s)
&=\frac{1}{|\mathcal S|K}
\sum_{(x,p,c,\mathbf y^*)\in\mathcal S} \\
&\quad\left\|g_s(x,p,c)-\mathbf y^*\right\|_2^2.
\end{aligned}
\label{eq:mse}
\end{equation}
This objective simultaneously constrains all five dimensions to align the predictions with expert ratings.

\subsection{The MuseCritic Reward Model}

\paragraph{Overview.}
To introduce an explicit textual intermediate representation into reward modeling, we treat natural-language aesthetic critiques as intermediate variables between a song and its final scores.
Given a song $x$ and rubric $p$, \method first generates
\begin{equation}
\hat{c}=g_c(x,p),
\label{eq:critic-generation}
\end{equation}
where $\hat c$ is a single critique with five labeled sections describing perceptible musical phenomena under the corresponding dimensions.
The shared backbone then receives the song, rubric, and self-generated critique to produce
\begin{equation}
\hat{\mathbf{y}}=g_s(x,p,\hat c).
\label{eq:critic-score}
\end{equation}
Relative to direct score prediction, Equation~\ref{eq:critic-score} introduces a variable-length textual representation before scoring and organizes descriptions of auditory evidence distributed across a long song according to the evaluation rubric.

\paragraph{Training MuseCritic.}
Ideally, \method would be trained on a dataset $\mathcal D^{*}=\{(x_i,p,c_i^{*},\mathbf y_i^{*})\}_{i=1}^{N}$, where $c_i^{*}$ is a human-authored aesthetic critique consistent with the corresponding expert ratings.
Such critiques would be authored by human music experts, but existing song datasets rarely pair expert scores with expert critiques.
Following prior work that constructs supervision from model feedback \citep{bai2022constitutional,ankner2024cloud}, we use the audio-capable Gemini-3-Pro \citep{google2025gemini3} to generate score-conditioned synthetic critiques.
Specifically, we begin with a score-only dataset $\mathcal D_{\mathrm{score}}=\{(x_i,\mathbf y_i^{*})\}_{i=1}^{N}$.
To generate a critique, Gemini observes the song $x_i$, rubric $p$, and expert mean ratings $\mathbf y_i^{*}$, then writes dimension-specific commentary whose evaluative polarity is consistent with each rating on a five-point Likert scale.
The prompt requests analysis of concrete musical phenomena while prohibiting numerical scores in the critique itself.
This procedure yields the offline critique dataset
\begin{equation}
\mathcal D_{\mathrm{off}}
=\{(x_i,p,c_i^{\mathrm{off}},\mathbf y_i^*)\}_{i=1}^{N}.
\label{eq:offline-data}
\end{equation}
Appendix~\ref{app:prompts} provides the complete external-teacher prompt.
To further ensure the reliability of the aesthetic critiques, we manually verify the generated critiques, as detailed in Appendix~\ref{app:human-review}.

Training comprises two stages.
In Stage I, we supervise the audio language model on $\mathcal D_{\mathrm{off}}$ with the critique-generation objective
\begin{equation}
\begin{aligned}
\mathcal{L}_{\mathrm{SFT}}
={}&-\frac{1}{\sum_{i=1}^{N}\lvert c_i^{\mathrm{off}}\rvert} \\
&\quad
\sum_{i=1}^{N}
\sum_{t=1}^{\lvert c_i^{\mathrm{off}}\rvert}
\log p_{\theta}\!\left(
c_{i,t}^{\mathrm{off}}
\mid x_i,p,c_{i,<t}^{\mathrm{off}}
\right).
\end{aligned}
\label{eq:sft}
\end{equation}
This is the standard next-token cross-entropy averaged over non-masked critique target tokens; the audio and user-prompt tokens do not contribute to the loss.
This stage produces a critique generator $g'_c$ with the ability to analyze a song along the five aesthetic dimensions.

In Stage II, we first use $g'_c$ to generate a new critique for every training example:
\begin{equation}
c_i^{\mathrm{on}}=g'_c(x_i,p).
\label{eq:online-critic}
\end{equation}
We replace the external-teacher critiques with these self-generated critiques and call the resulting collection the self-generated critique dataset $\mathcal D_{\mathrm{on}}$:
\begin{equation}
\mathcal D_{\mathrm{on}}
=\{(x_i,p,c_i^{\mathrm{on}},\mathbf y_i^*)\}_{i=1}^{N}.
\label{eq:online-replacement}
\end{equation}
We then initialize the shared \method backbone from the SFT model and jointly update the backbone LoRA parameters and all reward-head parameters on $\mathcal D_{\mathrm{on}}$.
The reward model thus learns from critiques generated by the model itself rather than by the external teacher, mitigating the critique distribution shift between reward training and inference.
The objective is the MSE from Equation~\ref{eq:mse} applied to $\mathcal D_{\mathrm{on}}$:
\begin{equation}
\mathcal L_{\mathrm{MC}}
=\mathcal L_{\mathrm{MSE}}(\mathcal D_{\mathrm{on}};g_s).
\label{eq:musecritic-mse}
\end{equation}
At inference time, the final LoRA-updated \method backbone generates an aesthetic critique according to Equation~\ref{eq:critic-generation} and then predicts five rewards according to Equation~\ref{eq:critic-score}.

\section{Experiments}
\label{sec:experiments}

\subsection{Experimental Setup}
\label{sec:setup}

\paragraph{Data.}
We use 2,199 songs from SongEval \citep{yao2025songeval} for training and reserve 200 songs for testing across all in-domain experiments; Appendix~\ref{app:data-details} provides the dataset and splitting details.

For out-of-domain evaluation, we use the 733 song-preference pairs from Music Arena \citep{kim2025musicarena}.
This dataset contains outputs from systems not represented in SongEval and provides pairwise human preferences rather than five-dimensional absolute scores.
It therefore measures preference generalization under distribution shift.

\paragraph{Baselines.}
The in-domain task requires five-dimensional song aesthetic scores.
We compare against two classes of baselines: (1) \textbf{Gemini-3.1-Pro} \citep{google2026gemini31}, an audio-capable LLM-as-a-Judge that generates critiques and five-dimensional ratings under the same rubric using the prompt in Figure~\ref{fig:judge-prompt}; and (2) \textbf{SongEval (UTMOS)}, a discriminative reward model that directly regresses five scores from audio.
SongEval does not release its original training code or data split, so we reimplement its published method and UTMOS architecture and train it for ten epochs on our 2,199/200 split, controlling for differences in training data.

Music Arena requires pairwise preference prediction.
In addition to \method, we evaluate the weights released by SongEval \citep{yao2025songeval}, the general-purpose audio aesthetic model Audiobox Aesthetics \citep{tjandra2025audiobox}, the open-source omni-modal model Qwen3-Omni-30B-A3B-Instruct \citep{xu2025qwen3omni}, the seven-dimensional reward model trained by SongBench \citep{wu2026songbench}, the scalar preference model TuneJury \citep{kim2026tunejury}, and CMI-RM \citep{ma2026cmirewardbench}.

\paragraph{Metrics.}
For in-domain evaluation, we report dimension-wise mean squared error (MSE), Pearson's linear correlation coefficient (LCC), Spearman's rank correlation coefficient (SRCC), and Kendall's $\tau$ (KTAU) between model predictions and expert ratings.
MSE measures agreement with the experts' absolute rating scale, while the three correlation coefficients measure relative agreement across songs in terms of linear association and ranking.
Each metric is computed independently over the 200 test songs for each dimension.

For out-of-domain evaluation, we report pairwise accuracy.
A prediction is correct only when the model assigns a strictly higher overall reward to the human-preferred song; ties count as errors.
For models that output multiple song-level scores, namely SongEval, Audiobox Aesthetics, \method, Qwen3-Omni, and SongBench, we average their respective dimensions with equal weight to obtain the final reward.
TuneJury's scalar preference score and CMI-RM's musicality score are used directly.
Qwen3-Omni uses the same rubric and generative scoring format as the in-domain Gemini baseline (Figure~\ref{fig:judge-prompt}).
The generative judge evaluations use deterministic decoding with $T=0$.

For downstream evaluation after GRPO, music quality is assessed using SongEval \citep{yao2025songeval} in terms of overall coherence (CO), memorability (ME), naturalness of vocal breathing and phrasing (NA), clarity of song structure (CL), and overall musicality (MU).
Audio quality is evaluated with Audiobox Aesthetics \citep{tjandra2025audiobox}, which considers content enjoyment (CE), content usefulness (CU), production complexity (PC), and production quality (PQ).

\begin{table*}[t]
\centering
\small
\setlength{\tabcolsep}{5pt}
\begin{tabular}{llrrrr}
\toprule
Dimension & Model & MSE $\downarrow$ & LCC $\uparrow$ & SRCC $\uparrow$ & KTAU $\uparrow$ \\
\midrule
\multirow{3}{*}{Coherence}
 & \textbf{\method} & \textbf{0.2058} & \textbf{0.9147} & \textbf{0.8894} & \textbf{0.7262} \\
 & Gemini-3.1-Pro & 0.9881 & 0.5906 & 0.5898 & 0.4359 \\
 & SongEval (UTMOS) & 0.2605 & 0.8876 & 0.8605 & 0.6835 \\
\midrule
\multirow{3}{*}{Memorability}
 & \textbf{\method} & \textbf{0.2511} & \textbf{0.9098} & \textbf{0.8839} & \textbf{0.7232} \\
 & Gemini-3.1-Pro & 1.1183 & 0.5806 & 0.5623 & 0.4113 \\
 & SongEval (UTMOS) & 0.3055 & 0.8857 & 0.8595 & 0.6837 \\
\midrule
\multirow{3}{*}{Vocal naturalness}
 & \textbf{\method} & \textbf{0.2537} & \textbf{0.8906} & \textbf{0.8678} & \textbf{0.6945} \\
 & Gemini-3.1-Pro & 1.9736 & 0.4921 & 0.5237 & 0.3877 \\
 & SongEval (UTMOS) & 0.3071 & 0.8593 & 0.8293 & 0.6491 \\
\midrule
\multirow{3}{*}{Structural clarity}
 & \textbf{\method} & \textbf{0.2275} & \textbf{0.9068} & \textbf{0.8806} & \textbf{0.7109} \\
 & Gemini-3.1-Pro & 1.4740 & 0.5532 & 0.5752 & 0.4162 \\
 & SongEval (UTMOS) & 0.2730 & 0.8831 & 0.8565 & 0.6790 \\
\midrule
\multirow{3}{*}{Musicality}
 & \textbf{\method} & \textbf{0.2202} & \textbf{0.9120} & \textbf{0.8972} & \textbf{0.7344} \\
 & Gemini-3.1-Pro & 0.9763 & 0.6152 & 0.6099 & 0.4459 \\
 & SongEval (UTMOS) & 0.2913 & 0.8807 & 0.8599 & 0.6876 \\
\bottomrule
\end{tabular}
\caption{\textbf{In-domain five-dimensional aesthetic scoring on the SongEval test set.} All models are evaluated on the fixed test set of 200 songs. SongEval (UTMOS) denotes the regressor retrained on the same 2,199/200 split. MSE measures absolute error against expert ratings; LCC, SRCC, and KTAU denote Pearson, Spearman, and Kendall $\tau$ correlations. Arrows indicate the preferred direction, and bold marks the best value for each metric and dimension.}
\label{tab:in-domain}
\end{table*}

\paragraph{Training.}
All \method variants use MOSS-Audio-8B-Instruct \citep{yang2026mossaudio} as the base model.
During critique cold-start, we fine-tune all backbone parameters for one epoch with a learning rate of $5\times10^{-5}$ to obtain the SFT-Critique-Generator.
We use the SFT-Critique-Generator to initialize \method and train it for ten epochs on the self-generated critique data.
The backbone uses LoRA (rank 8, scaling factor 32) \citep{hu2021lora}, while the reward head is fully updated with a learning rate of $2\times10^{-4}$.
Appendix~\ref{app:training-details} provides the complete optimization configuration.

\subsection{Evaluation Results}
\label{sec:main-results}

\subsubsection{In-Domain Song Aesthetic Scoring}

Table~\ref{tab:in-domain} compares model predictions with the expert mean ratings of the 200-song held-out SongEval test set.
\method achieves the lowest MSE and highest correlation on all five dimensions.
Relative to the SongEval (UTMOS) regression baseline trained on the same data, \method reduces macro-averaged MSE from 0.2875 to 0.2316.
Macro-averaged LCC, SRCC, and KTAU increase from 0.8793, 0.8531, and 0.6766 to 0.9068, 0.8838, and 0.7178.
Section~\ref{sec:ablation} isolates the effect of critique modeling through controlled ablations.

Although Gemini-3.1-Pro has not been specialized for song aesthetics, it generates coherent critiques.
Its macro-average MSE of 1.3061, however, substantially exceeds the 0.2316 of \method, and its dimension-wise correlations fall below those of models supervised by expert scores.
This result indicates that an unadapted LLM-as-a-Judge cannot reliably reproduce five-dimensional expert judgments of song aesthetics.
By supervising the reward head with expert ratings, \method maps information in the critiques to continuous scores that better align with expert assessments.

\begin{table}[t]
\centering
\small
\begin{tabular}{lr}
\toprule
Model & Accuracy (\%) $\uparrow$ \\
\midrule
Qwen3-Omni-30B-A3B & 53.75 \\
Audiobox Aesthetics & 68.49 \\
TuneJury & 67.53 \\
SongEval & 70.80 \\
SongBench & 71.08 \\
CMI-RM & 71.21 \\
\textbf{\method} & \textbf{71.35} \\
\bottomrule
\end{tabular}
\caption{\textbf{Out-of-domain pairwise preference accuracy on Music Arena.} A prediction is correct when the model assigns a higher overall reward to the human-preferred song; ties count as errors.}
\label{tab:arena}
\end{table}

\subsubsection{Out-of-Domain Generalization}
We evaluate whether the five-dimensional aesthetic scores learned from SongEval transfer to pairwise human-preference prediction on Music Arena, which contains songs from generation systems absent from the training dataset.
As shown in Table~\ref{tab:arena}, \method achieves 71.35\% accuracy and is competitive with CMI-RM (71.21\%) and SongBench (71.08\%).
These results indicate that the learned scores remain informative for human-preference prediction under this distribution shift.
Together with the in-domain results, this evaluation shows that \method supports both expert-referenced multidimensional scoring and out-of-domain pairwise preference prediction.

\subsection{Downstream Reinforcement Learning}
\label{sec:downstream}

To evaluate the utility of \method as an optimization signal, we train Muse-0.6B \citep{jiang2026muse} with GRPO \citep{shao2024deepseekmath}.
We additionally train a control policy using the w/o critique reward model, which shares the same critique-SFT initialization as \method but omits critiques during reward learning and scoring.
We denote the resulting policies as \musedirect and \musegrpo, respectively.
Using seed 42, we sample 500 multi-turn song-generation examples from the public Muse training set while preserving the original language ratio: 195 Chinese and 305 English examples.
For either reward model, the reinforcement-learning reward for a generated song $\tilde x$ is the mean of its five predicted aesthetic scores:
\begin{equation}
r(\tilde x)=\frac{1}{K}\sum_{k=1}^{K}\hat y_k(\tilde x),\qquad K=5.
\end{equation}
Both policies use the same hyperparameters: one epoch on six H200 GPUs, eight sampled outputs per prompt, and a learning rate of $10^{-6}$.
Appendix~\ref{app:grpo-details} gives the complete rollout configuration.

Evaluation uses the 100 multi-turn test prompts from Muse \citep{jiang2026muse}.
All three policies use greedy decoding with $T=0$.
Appendix~\ref{app:grpo-details} describes the complete evaluation protocol and robustness checks across evaluation windows.
As shown in Table~\ref{tab:downstream}, \musegrpo obtains higher observed values than Muse on all nine Audiobox Aesthetics and SongEval metrics.
The largest Audiobox gains occur in production complexity (PC, $+0.15$) and content enjoyment (CE, $+0.12$), while all five SongEval dimensions improve by $0.05$--$0.06$.
Moreover, \musegrpo outperforms \musedirect on every metric.
These results support the downstream utility of \method and show that critique-conditioned rewards provide additional gains over direct reward modeling.

\begin{table*}[t]
\centering
\small
\setlength{\tabcolsep}{3.6pt}
\begin{tabular}{lrrrrrrrrr}
\toprule
& \multicolumn{4}{c}{Audiobox Aesthetics $\uparrow$} & \multicolumn{5}{c}{SongEval $\uparrow$} \\
\cmidrule(lr){2-5}\cmidrule(lr){6-10}
Model & CE & CU & PC & PQ & CO & MU & ME & CL & NA \\
\midrule
Muse & 7.33 & 7.64 & 6.37 & 8.11 & 4.01 & 3.84 & 3.94 & 3.89 & 3.83 \\
\musedirect & 7.38 & 7.65 & 6.44 & 8.10 & 4.05 & 3.87 & 3.99 & 3.92 & 3.86 \\
\musegrpo & \textbf{7.45} & \textbf{7.66} & \textbf{6.52} & \textbf{8.12} & \textbf{4.07} & \textbf{3.89} & \textbf{4.00} & \textbf{3.94} & \textbf{3.89} \\
\bottomrule
\end{tabular}
\caption{\textbf{Downstream aesthetic evaluation of Muse-0.6B policies optimized with GRPO.}
\musedirect uses the no-critique reward model, whereas \musegrpo uses the full critique-conditioned reward model.
All three policies are evaluated on 100 multi-turn Muse prompts with $T=0$ decoding using Audiobox Aesthetics (CE/CU/PC/PQ) and SongEval (CO/MU/ME/CL/NA).
Arrows indicate the preferred direction, and bold marks the best result.}
\label{tab:downstream}
\end{table*}

\begin{figure*}[t]
    \centering
    \includegraphics[width=0.9\textwidth]{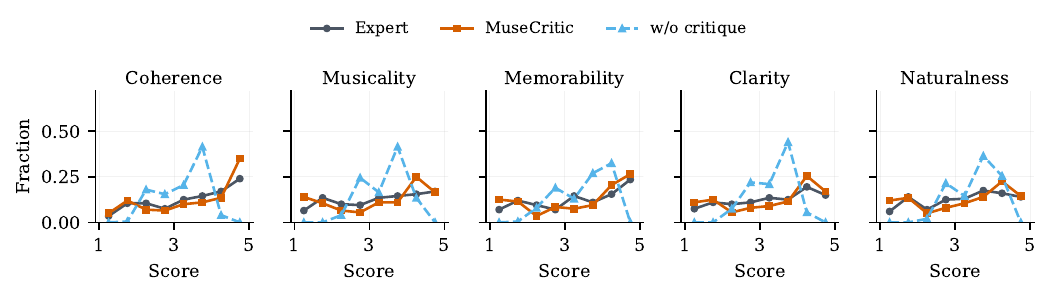}
    \caption{\textbf{Five-dimensional score distributions on the SongEval test set.} Curves compare expert ratings with predictions from \method and its no-critique variant. Without critiques, predictions concentrate in a narrow high-score range, whereas \method more closely matches the expert distributions.}
    \label{fig:score-distributions}
\end{figure*}

\subsection{Ablation Studies}
\label{sec:ablation}

Table~\ref{tab:ablation} studies three design choices across all five dimensions: natural-language aesthetic critiques, self-generated critiques, and SFT initialization before reward learning.
All variants use the same training songs, reward head, and optimization configuration.

\begin{table*}[t]
\centering
\small
\setlength{\tabcolsep}{4.5pt}
\begin{tabular}{llrrrr}
\toprule
Dimension & Model variant & MSE $\downarrow$ & LCC $\uparrow$ & SRCC $\uparrow$ & KTAU $\uparrow$ \\
\midrule
\multirow{4}{*}{Coherence}
 & \method & \textbf{0.2058} & 0.9147 & 0.8894 & 0.7262 \\
 & w/o critique & 0.3881 & \textbf{0.9173} & \textbf{0.8954} & \textbf{0.7326} \\
 & Offline critique & 0.4821 & 0.8342 & 0.7566 & 0.5675 \\
 & w/o SFT & 0.8556 & 0.6300 & 0.6066 & 0.4689 \\
\midrule
\multirow{4}{*}{Memorability}
 & \method & \textbf{0.2511} & \textbf{0.9098} & 0.8839 & \textbf{0.7232} \\
 & w/o critique & 0.4936 & 0.9048 & \textbf{0.8842} & 0.7219 \\
 & Offline critique & 0.6635 & 0.8141 & 0.7590 & 0.5672 \\
 & w/o SFT & 1.0037 & 0.6175 & 0.6378 & 0.4823 \\
\midrule
\multirow{4}{*}{Vocal naturalness}
 & \method & \textbf{0.2537} & \textbf{0.8906} & 0.8678 & 0.6945 \\
 & w/o critique & 0.6186 & 0.8899 & \textbf{0.8684} & \textbf{0.7034} \\
 & Offline critique & 0.5739 & 0.7936 & 0.7527 & 0.5674 \\
 & w/o SFT & 0.6469 & 0.6779 & 0.6768 & 0.4978 \\
\midrule
\multirow{4}{*}{Structural clarity}
 & \method & \textbf{0.2275} & \textbf{0.9068} & \textbf{0.8806} & \textbf{0.7109} \\
 & w/o critique & 0.4635 & 0.9041 & 0.8785 & 0.7088 \\
 & Offline critique & 0.5033 & 0.8145 & 0.7479 & 0.5526 \\
 & w/o SFT & 0.6681 & 0.6869 & 0.6764 & 0.5061 \\
\midrule
\multirow{4}{*}{Musicality}
 & \method & \textbf{0.2202} & \textbf{0.9120} & \textbf{0.8972} & \textbf{0.7344} \\
 & w/o critique & 0.5386 & 0.9080 & 0.8875 & 0.7293 \\
 & Offline critique & 0.5175 & 0.8205 & 0.7765 & 0.5877 \\
 & w/o SFT & 0.5960 & 0.7354 & 0.7325 & 0.5488 \\
\bottomrule
\end{tabular}
\caption{\textbf{In-domain ablations of the key \method training choices.} All variants use the same 2,199/200 SongEval split. ``w/o critique'' predicts scores directly, ``Offline critique'' trains on Gemini critiques, and ``w/o SFT'' generates training critiques with the unadapted MOSS-Audio model. The full model combines SFT initialization with self-generated critiques. Arrows indicate the preferred direction, and bold marks the best value for each metric and dimension.}
\label{tab:ablation}
\end{table*}

\paragraph{Self-generated aesthetic critiques improve absolute-score agreement.}
Removing the critiques increases macro-average MSE from 0.2316 to 0.5005, while the macro-averaged rank correlations remain close.
Figure~\ref{fig:score-distributions} explains this difference: the no-critique model concentrates its predictions in a relatively narrow high-score region, whereas \method produces a score distribution closer to the expert ratings.
A reward model should preserve not only rank order but also the relative magnitude of quality differences through score intervals.
An overly contracted prediction distribution reduces agreement with experts' absolute ratings and weakens reward discrimination among samples of different quality during downstream policy optimization.
With critiques, evaluative polarity and concrete musical descriptions provide an explicit semantic reference, bringing the prediction distribution closer to the expert scale and reducing MSE in every dimension.

\paragraph{Both audio and matched critiques contribute to scoring.}
To further analyze the roles of audio and critiques in score prediction, we apply two inference-time interventions to the fixed \method model: randomly mismatching critiques across test songs and removing the audio input at the scoring stage while retaining each song's corresponding critique.
We repeat random mismatch with five independent permutations; Appendix~\ref{app:intervention-details} reports the complete results.
Both interventions increase MSE and reduce correlation on all five dimensions.
The degradation under mismatched critiques shows that song-specific critiques contribute materially to score prediction rather than serving only as accompanying textual explanations.
Performance also declines when the scoring stage receives the matched critique but not the audio, indicating that the critique does not fully replace audio information in the current model.

\paragraph{Training on self-generated critiques improves reward prediction.}
To isolate the role of the self-generated dataset $\mathcal D_{\mathrm{on}}$, we train a control model on $\mathcal D_{\mathrm{off}}$, using external-teacher critiques instead of critiques generated by the target model.
As Table~\ref{tab:ablation} shows, the full \method trained on $\mathcal D_{\mathrm{on}}$ achieves higher LCC, SRCC, and KTAU than the offline-critique variant on every dimension.
Replacing $\mathcal D_{\mathrm{on}}$ with $\mathcal D_{\mathrm{off}}$ increases macro-average MSE from 0.2316 to 0.5481 and decreases macro-average LCC, SRCC, and KTAU from 0.9068, 0.8838, and 0.7178 to 0.8154, 0.7585, and 0.5685, respectively.
We suggest that the external teacher and MOSS-Audio differ in vocabulary, information density, and patterns.
Training only on teacher critiques therefore creates a larger mismatch between the critiques observed during reward training and those generated at inference time.

\paragraph{SFT initializes effective critique generation.}
Without SFT, we use the unadapted MOSS-Audio-8B-Instruct model to generate critiques and subsequently train the reward model on these critiques. This variant performs worst overall: macro-averaged MSE increases from 0.2316 to 0.7541, while macro-averaged LCC, SRCC, and KTAU decrease from 0.9068, 0.8838, and 0.7178 to 0.6695, 0.6660, and 0.5008, respectively. The degradation in both absolute-score accuracy and ranking consistency indicates that self-generation alone does not produce a reliable intermediate representation for reward learning. Supervised fine-tuning on teacher critiques is therefore necessary for the critique generator to acquire the evaluation criteria, domain-specific terminology, and rating semantics required to distinguish both the level and relative ordering of song quality.

\section{Conclusion}
\label{sec:conclusion}
We introduced \method, a semi-scalar reward model that uses natural-language aesthetic critiques as intermediate representations for five-dimensional evaluation of complete songs.
\method learns critique generation from external-teacher supervision and trains its reward head on self-generated critiques, reducing scoring error while maintaining strong rank correlations on SongEval and generalizing to Music Arena.
When used for GRPO, \method improves Muse across all nine SongEval and Audiobox Aesthetics metrics, demonstrating its utility as an optimization signal for song generation.

\section*{Limitations}

The critique-then-score procedure requires autoregressive critique generation before reward prediction and is therefore more computationally expensive than direct score regression.
Moreover, long-form song datasets that pair audio with expert aesthetic ratings remain scarce.
Consequently, \method is trained primarily on the Chinese and English vocal songs in SongEval.
Future work can extend training to a broader range of languages and musical genres.

\phantomsection\label{page:references-start}
\bibliography{main}

@article{yao2025songeval,
  title = {{SongEval}: A Benchmark Dataset for Song Aesthetics Evaluation},
  author = {Jixun Yao and Guobin Ma and Huixin Xue and Huakang Chen and Chunbo Hao and Yuepeng Jiang and Haohe Liu and Ruibin Yuan and Jin Xu and Wei Xue and Hao Liu and Lei Xie},
  journal = {arXiv preprint arXiv:2505.10793},
  year = {2025},
  url = {https://arxiv.org/abs/2505.10793}
}

@article{yang2026mossaudio,
  title = {{MOSS-Audio} Technical Report},
  author = {Chen Yang and Chufan Yu and Hanfu Chen and Jie Zhu and Jingqi Chen and Ke Chen and Wenxuan Wang and Yang Wang and Yaozhou Jiang and Yi Jiang and Zhengyuan Lin and Ziqi Chen and Zhaoye Fei and Chenghao Liu and Donghua Yu and Jun Zhan and Kang Yu and Kexin Huang and Liwei Fan and Mingshu Chen and Qinyuan Cheng and Ruixiao Li and Shimin Li and Songlin Wang and Xingjian Zhao and Yang Gao and Yitian Gong and Yiyang Zhang and Zhe Xu and Xipeng Qiu},
  journal = {arXiv preprint arXiv:2606.01802},
  year = {2026},
  url = {https://arxiv.org/abs/2606.01802}
}

@inproceedings{jiang2026muse,
  title = {Muse: Towards Reproducible Long-Form Song Generation with Fine-Grained Style Control},
  author = {Changhao Jiang and Jiahao Chen and Zhenghao Xiang and Zhixiong Yang and Hanchen Wang and Jiabao Zhuang and Xinmeng Che and Jiajun Sun and Hui Li and Yifei Cao and Shihan Dou and Ming Zhang and Junjie Ye and Tao Ji and Tao Gui and Qi Zhang and Xuanjing Huang},
  booktitle = {Findings of the Association for Computational Linguistics: ACL 2026},
  pages = {22492--22512},
  year = {2026},
  url = {https://aclanthology.org/2026.findings-acl.1129/}
}

@inproceedings{kim2025musicarena,
  title = {Music Arena: Live Evaluation for Text-to-Music},
  author = {Yonghyun Kim and Wayne Chi and Anastasios N. Angelopoulos and Wei-Lin Chiang and Koichi Saito and Shinji Watanabe and Yuki Mitsufuji and Chris Donahue},
  booktitle = {Advances in Neural Information Processing Systems 38, Creative AI Track},
  year = {2025},
  url = {https://arxiv.org/abs/2507.20900}
}

@article{tjandra2025audiobox,
  title = {Meta {Audiobox Aesthetics}: Unified Automatic Quality Assessment for Speech, Music, and Sound},
  author = {Andros Tjandra and Yi-Chiao Wu and Baishan Guo and John Hoffman and Brian Ellis and Apoorv Vyas and Bowen Shi and Sanyuan Chen and Matt Le and Nick Zacharov and Carleigh Wood and Ann Lee and Wei-Ning Hsu},
  journal = {arXiv preprint arXiv:2502.05139},
  year = {2025},
  url = {https://arxiv.org/abs/2502.05139}
}

@article{xu2025qwen3omni,
  title = {{Qwen3-Omni} Technical Report},
  author = {Jin Xu and Zhifang Guo and Hangrui Hu and Yunfei Chu and Xiong Wang and Jinzheng He and Yuxuan Wang and Xian Shi and Ting He and Xinfa Zhu and Yuanjun Lv and Yongqi Wang and Dake Guo and He Wang and Linhan Ma and Pei Zhang and Xinyu Zhang and Hongkun Hao and Zishan Guo and Baosong Yang and Bin Zhang and Ziyang Ma and Xipin Wei and Shuai Bai and Keqin Chen and Xuejing Liu and Peng Wang and Mingkun Yang and Dayiheng Liu and Xingzhang Ren and Bo Zheng and Rui Men and Fan Zhou and Bowen Yu and Jianxin Yang and Le Yu and Jingren Zhou and Junyang Lin},
  journal = {arXiv preprint arXiv:2509.17765},
  year = {2025},
  url = {https://arxiv.org/abs/2509.17765}
}

@article{wu2026songbench,
  title = {{SongBench}: A Fine-Grained Multi-Aspect Benchmark for Song Quality Assessment},
  author = {Dapeng Wu and Shun Lei and Wei Tan and Guangzheng Li and Yunzhe Wang and Huaicheng Zhang and Lishi Zuo and Zhiyong Wu},
  journal = {arXiv preprint arXiv:2604.25937},
  year = {2026},
  doi = {10.48550/arXiv.2604.25937},
  url = {https://arxiv.org/abs/2604.25937}
}

@article{kim2026tunejury,
  title = {{TuneJury}: An Open Metric for Improving Music Generation Preference Alignment},
  author = {Yonghyun Kim and Junwon Lee and Haiwen Xia and Yinghao Ma and Junghyun Koo and Koichi Saito and Yuki Mitsufuji and Chris Donahue},
  journal = {arXiv preprint arXiv:2606.17006},
  year = {2026},
  doi = {10.48550/arXiv.2606.17006},
  url = {https://arxiv.org/abs/2606.17006}
}

@article{ma2026cmirewardbench,
  title = {{CMI-RewardBench}: Evaluating Music Reward Models with Compositional Multimodal Instruction},
  author = {Yinghao Ma and Haiwen Xia and Hewei Gao and Weixiong Chen and Yuxin Ye and Yuchen Yang and Sungkyun Chang and Mingshuo Ding and Yizhi Li and Ruibin Yuan and Simon Dixon and Emmanouil Benetos},
  journal = {arXiv preprint arXiv:2603.00610},
  year = {2026},
  doi = {10.48550/arXiv.2603.00610},
  url = {https://arxiv.org/abs/2603.00610}
}

@article{shao2024deepseekmath,
  title = {{DeepSeekMath}: Pushing the Limits of Mathematical Reasoning in Open Language Models},
  author = {Zhihong Shao and Peiyi Wang and Qihao Zhu and Runxin Xu and Junxiao Song and Xiao Bi and Haowei Zhang and Mingchuan Zhang and Y. K. Li and Y. Wu and Daya Guo},
  journal = {arXiv preprint arXiv:2402.03300},
  year = {2024},
  url = {https://arxiv.org/abs/2402.03300}
}

@article{cideron2024musicrl,
  title = {{MusicRL}: Aligning Music Generation to Human Preferences},
  author = {Geoffrey Cideron and Sertan Girgin and Mauro Verzetti and Damien Vincent and Matej Kastelic and Zal{\'a}n Borsos and Brian McWilliams and Victor Ungureanu and Olivier Bachem and Olivier Pietquin and Matthieu Geist and L{\'e}onard Hussenot and Neil Zeghidour and Andrea Agostinelli},
  journal = {arXiv preprint arXiv:2402.04229},
  year = {2024},
  url = {https://arxiv.org/abs/2402.04229}
}

@article{hu2021lora,
  title = {{LoRA}: Low-Rank Adaptation of Large Language Models},
  author = {Edward J. Hu and Yelong Shen and Phillip Wallis and Zeyuan Allen-Zhu and Yuanzhi Li and Shean Wang and Lu Wang and Weizhu Chen},
  journal = {arXiv preprint arXiv:2106.09685},
  year = {2021},
  url = {https://arxiv.org/abs/2106.09685}
}

@article{agostinelli2023musiclm,
  title = {{MusicLM}: Generating Music From Text},
  author = {Andrea Agostinelli and Timo I. Denk and Zal{\'a}n Borsos and Jesse Engel and Mauro Verzetti and Antoine Caillon and Qingqing Huang and Aren Jansen and Adam Roberts and Marco Tagliasacchi and Matt Sharifi and Neil Zeghidour and Christian Frank},
  journal = {arXiv preprint arXiv:2301.11325},
  year = {2023},
  url = {https://arxiv.org/abs/2301.11325}
}

@inproceedings{copet2023musicgen,
  title = {Simple and Controllable Music Generation},
  author = {Jade Copet and Felix Kreuk and Itai Gat and Tal Remez and David Kant and Gabriel Synnaeve and Yossi Adi and Alexandre D{\'e}fossez},
  booktitle = {Advances in Neural Information Processing Systems},
  year = {2023},
  url = {https://arxiv.org/abs/2306.05284}
}

@misc{google2025gemini3,
  author = {{Google}},
  title = {{Gemini 3 Pro Preview}: Model Documentation},
  year = {2025},
  howpublished = {Google AI for Developers},
  url = {https://ai.google.dev/gemini-api/docs/models/gemini-3-pro-preview},
  note = {Accessed 2026-08-05}
}

@misc{google2026gemini31,
  author = {{Google DeepMind}},
  title = {{Gemini 3.1 Pro} Model Card},
  year = {2026},
  url = {https://deepmind.google/models/model-cards/gemini-3-1-pro},
  note = {Published 2026-02-19; accessed 2026-08-05}
}

@article{ning2025diffrhythm,
  title = {{DiffRhythm}: Blazingly Fast and Embarrassingly Simple End-to-End Full-Length Song Generation with Latent Diffusion},
  author = {Ziqian Ning and Huakang Chen and Yuepeng Jiang and Chunbo Hao and Guobin Ma and Shuai Wang and Jixun Yao and Lei Xie},
  journal = {arXiv preprint arXiv:2503.01183},
  year = {2025},
  doi = {10.48550/arXiv.2503.01183},
  url = {https://arxiv.org/abs/2503.01183}
}

@article{yuan2025yue,
  title = {{YuE}: Scaling Open Foundation Models for Long-Form Music Generation},
  author = {Ruibin Yuan and Hanfeng Lin and Shuyue Guo and Ge Zhang and Jiahao Pan and Yongyi Zang and Haohe Liu and Yiming Liang and Wenye Ma and Xingjian Du and others},
  journal = {arXiv preprint arXiv:2503.08638},
  year = {2025},
  doi = {10.48550/arXiv.2503.08638},
  url = {https://arxiv.org/abs/2503.08638}
}

@article{lei2025levo,
  title = {{LeVo}: High-Quality Song Generation with Multi-Preference Alignment},
  author = {Shun Lei and Yaoxun Xu and Zhiwei Lin and Huaicheng Zhang and Wei Tan and Hangting Chen and Jianwei Yu and Yixuan Zhang and Chenyu Yang and Haina Zhu and Shuai Wang and Zhiyong Wu and Dong Yu},
  journal = {arXiv preprint arXiv:2506.07520},
  year = {2025},
  doi = {10.48550/arXiv.2506.07520},
  url = {https://arxiv.org/abs/2506.07520}
}

@article{gong2025acestep,
  title = {{ACE-Step}: A Step Towards Music Generation Foundation Model},
  author = {Junmin Gong and Sean Zhao and Sen Wang and Shengyuan Xu and Joe Guo},
  journal = {arXiv preprint arXiv:2506.00045},
  year = {2025},
  doi = {10.48550/arXiv.2506.00045},
  url = {https://arxiv.org/abs/2506.00045}
}

@article{huang2022mulan,
  title = {{MuLan}: A Joint Embedding of Music Audio and Natural Language},
  author = {Qingqing Huang and Aren Jansen and Joonseok Lee and Ravi Ganti and Judith Yue Li and Daniel P. W. Ellis},
  journal = {arXiv preprint arXiv:2208.12415},
  year = {2022},
  doi = {10.48550/arXiv.2208.12415},
  url = {https://arxiv.org/abs/2208.12415}
}

@article{zhu2025muq,
  title = {{MuQ}: Self-Supervised Music Representation Learning with Mel Residual Vector Quantization},
  author = {Haina Zhu and Yizhi Zhou and Hangting Chen and Jianwei Yu and Ziyang Ma and Rongzhi Gu and Yi Luo and Wei Tan and Xie Chen},
  journal = {arXiv preprint arXiv:2501.01108},
  year = {2025},
  doi = {10.48550/arXiv.2501.01108},
  url = {https://arxiv.org/abs/2501.01108}
}

@article{deshmukh2024pam,
  title = {{PAM}: Prompting Audio-Language Models for Audio Quality Assessment},
  author = {Soham Deshmukh and Dareen Alharthi and Benjamin Elizalde and Hannes Gamper and Mahmoud Al Ismail and Rita Singh and Bhiksha Raj and Huaming Wang},
  journal = {arXiv preprint arXiv:2402.00282},
  year = {2024},
  doi = {10.48550/arXiv.2402.00282},
  url = {https://arxiv.org/abs/2402.00282}
}

@article{liu2025musiceval,
  title = {{MusicEval}: A Generative Music Corpus with Expert Ratings for Automatic Text-to-Music Evaluation},
  author = {Cheng Liu and Hui Wang and Jinghua Zhao and Shiwan Zhao and Hui Bu and Xin Xu and Jiaming Zhou and Haoqin Sun and Yong Qin},
  journal = {arXiv preprint arXiv:2501.10811},
  year = {2025},
  doi = {10.48550/arXiv.2501.10811},
  url = {https://arxiv.org/abs/2501.10811}
}

@article{ouyang2022instructgpt,
  title = {Training Language Models to Follow Instructions with Human Feedback},
  author = {Long Ouyang and Jeff Wu and Xu Jiang and Diogo Almeida and Carroll L. Wainwright and Pamela Mishkin and Chong Zhang and Sandhini Agarwal and Katarina Slama and Alex Ray and John Schulman and Jacob Hilton and Fraser Kelton and Luke Miller and Maddie Simens and Amanda Askell and Peter Welinder and Paul Christiano and Jan Leike and Ryan Lowe},
  journal = {arXiv preprint arXiv:2203.02155},
  year = {2022},
  doi = {10.48550/arXiv.2203.02155},
  url = {https://arxiv.org/abs/2203.02155}
}

@article{zheng2023llmjudge,
  title = {Judging {LLM}-as-a-Judge with {MT-Bench} and Chatbot Arena},
  author = {Lianmin Zheng and Wei-Lin Chiang and Ying Sheng and Siyuan Zhuang and Zhanghao Wu and Yonghao Zhuang and Zi Lin and Zhuohan Li and Dacheng Li and Eric P. Xing and Hao Zhang and Joseph E. Gonzalez and Ion Stoica},
  journal = {arXiv preprint arXiv:2306.05685},
  year = {2023},
  doi = {10.48550/arXiv.2306.05685},
  url = {https://arxiv.org/abs/2306.05685}
}

@article{bai2022constitutional,
  title = {Constitutional {AI}: Harmlessness from {AI} Feedback},
  author = {Yuntao Bai and Saurav Kadavath and Sandipan Kundu and Amanda Askell and Jackson Kernion and Andy Jones and Anna Chen and Anna Goldie and Azalia Mirhoseini and Cameron McKinnon and others},
  journal = {arXiv preprint arXiv:2212.08073},
  year = {2022},
  doi = {10.48550/arXiv.2212.08073},
  url = {https://arxiv.org/abs/2212.08073}
}

@article{ankner2024cloud,
  title = {Critique-out-Loud Reward Models},
  author = {Zachary Ankner and Mansheej Paul and Brandon Cui and Jonathan D. Chang and Prithviraj Ammanabrolu},
  journal = {arXiv preprint arXiv:2408.11791},
  year = {2024},
  doi = {10.48550/arXiv.2408.11791},
  url = {https://arxiv.org/abs/2408.11791}
}

@inproceedings{yu2025criticrm,
  title = {Self-Generated Critiques Boost Reward Modeling for Language Models},
  author = {Yue Yu and Zhengxing Chen and Aston Zhang and Liang Tan and Chenguang Zhu and Richard Yuanzhe Pang and Yundi Qian and Xuewei Wang and Suchin Gururangan and Chao Zhang and Melanie Kambadur and Dhruv Mahajan and Rui Hou},
  booktitle = {Proceedings of the 2025 Conference of the Nations of the Americas Chapter of the Association for Computational Linguistics: Human Language Technologies (Volume 1: Long Papers)},
  pages = {11499--11514},
  publisher = {Association for Computational Linguistics},
  year = {2025},
  doi = {10.18653/v1/2025.naacl-long.573},
  url = {https://aclanthology.org/2025.naacl-long.573/}
}

@article{zhang2024genrm,
  title = {Generative Verifiers: Reward Modeling as Next-Token Prediction},
  author = {Lunjun Zhang and Arian Hosseini and Hritik Bansal and Mehran Kazemi and Aviral Kumar and Rishabh Agarwal},
  journal = {arXiv preprint arXiv:2408.15240},
  year = {2024},
  doi = {10.48550/arXiv.2408.15240},
  url = {https://arxiv.org/abs/2408.15240}
}

@article{liu2025deepseekgrm,
  title = {Inference-Time Scaling for Generalist Reward Modeling},
  author = {Zijun Liu and Peiyi Wang and Runxin Xu and Shirong Ma and Chong Ruan and Peng Li and Yang Liu and Yu Wu},
  journal = {arXiv preprint arXiv:2504.02495},
  year = {2025},
  doi = {10.48550/arXiv.2504.02495},
  url = {https://arxiv.org/abs/2504.02495}
}

\appendix
\section{Data Construction and Human Quality Review}
\label{app:data-details}

\subsection{Training Data Construction and Splitting}

\paragraph{SongEval split.}
SongEval contains 2,399 complete Chinese and English songs with five-dimensional aesthetic scores provided by annotators with music expertise.
Because the original work does not release a fixed train--test split, we load all 2,399 examples and shuffle them once using random seed 42.
The first 2,199 examples form the training set, and the final 200 examples form the test set.
We construct the split before generating critiques and reuse the same test examples across all main experiments, baselines, and ablation studies.
Consequently, no test song appears in either the SFT data or the reward-learning data.

\paragraph{Personally identifying information and offensive content.}
We inspect the data used in this study and confirm that it contains no personally identifying information or offensive content.

\paragraph{Data access and use.}
SongEval is a publicly released dataset, and we use it solely for academic research in accordance with its stated terms of use.

\paragraph{Offline critique generation.}
The original SongEval examples contain only song audio and five-dimensional mean expert ratings.
To construct $\mathcal D_{\mathrm{off}}$, we provide Gemini-3-Pro with the song, definitions of the five dimensions, and the expert mean ratings.
The prompt asks the model to infer perceptible musical evidence consistent with the scores on a five-point Likert scale, including melodic repetition and variation, section-level energy, rhythmic density, harmonic progression, vocal breathing, and changes in arrangement.
The critique must neither state numerical scores nor invent the identity of a singer or band.
We retain the generated aesthetic critique together with the original expert ratings.

\paragraph{Self-generated critique construction.}
After the offline critiques pass the human review described below and SFT is complete, we run the resulting SFT-Critique-Generator over all 2,199 training songs.
We use \texttt{do\_sample=False}, \texttt{num\_beams=1}, and at most 4,096 new tokens, corresponding to greedy decoding with $T=0$.
We replace each teacher critique with the corresponding self-generated critique while preserving the audio path, evaluation prompt, and expert mean ratings, thereby obtaining $\mathcal D_{\mathrm{on}}$.

\subsection{Human Quality Review}
\label{app:human-review}

\paragraph{Human quality review of LLM-generated critiques.}
After critique generation and before SFT, we recruit eight experts with music-domain expertise to manually review the critiques generated by Gemini-3-Pro.
Each critique is reviewed by two experts, who listen to the corresponding song and assess whether the critique agrees with human perceptual judgments and is grounded in audible musical evidence.
Factual accuracy and the absence of unsupported or hallucinated claims serve as the primary acceptance criteria.
The review also considers analytical depth, whether the claims refer to perceptible musical phenomena, coverage of all five evaluation dimensions, logical coherence, and clarity of expression.
A critique is accepted only when both reviewers judge that it satisfies these criteria.
For rejected critiques, we prompt Gemini-3-Pro to revise and regenerate the critique and submit the revision to another round of review.
Only accepted critiques are retained as SFT examples in $\mathcal D_{\mathrm{off}}$, yielding a final offline dataset of 2,199 examples.

\paragraph{Human quality evaluation of MuseCritic-generated critiques.}
To assess the reliability of critiques generated by the trained \method model, we randomly sample 100 of the 200 songs in the in-domain test set for human evaluation.
Six experts with music-domain expertise participate, and each critique is independently reviewed by three experts using the core quality criteria defined above.
We count a critique as passing by majority vote when at least two reviewers accept it.
Among the 100 critiques, 82 are accepted by all three reviewers, 14 by two reviewers, and 4 by one reviewer.
The resulting majority-vote pass rate is 96.0\% (Wilson 95\% confidence interval: 90.16\%--98.43\%), and the unanimous pass rate is 82.0\%.
These results show that most model-generated critiques in the sampled test set satisfy the specified human quality criteria.

\section{Gemini Model Versions}
\label{app:gemini-versions}

The external teacher referred to as Gemini-3-Pro in the main text uses the model identifier gemini-3-pro-preview.
The in-domain LLM-as-a-Judge baseline referred to as Gemini-3.1-Pro uses gemini-3.1-pro-preview.

\section{Complete Training Configuration}
\label{app:training-details}

\paragraph{Critique supervised fine-tuning.}
We fully fine-tune MOSS-Audio-8B-Instruct on $\mathcal D_{\mathrm{off}}$ using four NVIDIA H200 GPUs, bfloat16 precision, a per-device batch size of 1, and gradient accumulation over 8 steps, giving a global batch size of 32.
The learning rate is $5\times10^{-5}$ with no weight decay, a 5\% warmup, cosine scheduling, a maximum sequence length of 10,000, and random seed 42.
We train for one epoch to obtain the SFT-Critique-Generator and use it to initialize both the critique generator and the reward-model backbone.

\paragraph{MuseCritic reward learning.}
Starting from the SFT-Critique-Generator, we train \method for ten epochs on $\mathcal D_{\mathrm{on}}$ using four NVIDIA H200 GPUs and bfloat16 precision.
The backbone is updated with LoRA of rank 8 and scaling factor 32, while the reward head is fully trainable with dropout probability 0.1.
We use a learning rate of $2\times10^{-4}$, weight decay of 0.1, a per-device batch size of 1, gradient accumulation over 8 steps, a global batch size of 32, a 5\% warmup, cosine scheduling, a maximum sequence length of 10,000, and random seed 42.

\paragraph{SongEval baseline training.}
The in-domain SongEval (UTMOS-based) baseline is fully trained for ten epochs on two NVIDIA H200 GPUs using bfloat16 precision, a learning rate of $1\times10^{-4}$, weight decay of 0.1, a per-device batch size of 1, gradient accumulation over 8 steps, and a global batch size of 16.
It uses the same 5\% warmup, cosine scheduling, and random seed 42.

\paragraph{Reward-head implementation.}
The final valid-token state from the last backbone layer first passes through dropout with probability 0.1, followed by a linear projection from $\mathbb R^d$ to $\mathbb R^5$ and a sigmoid range transformation that produces scores in $(1,5)^5$.
Reward learning uses DeepSpeed ZeRO-3.
LoRA is applied to the language backbone, while the reward head remains fully trainable.
Training uses only the MSE objective in Equation~\ref{eq:musecritic-mse}, without an auxiliary critique language-modeling loss.

\paragraph{Checkpoints and inference.}
We train for ten epochs to obtain the final \method model used in the main experiments.
During in-domain testing, the model generates at most 4,096 new tokens using greedy decoding with $T=0$.
The five rewards are computed from the hidden state of the final valid token---the end-of-turn token following the critique---in the complete song--rubric--self-generated-critique sequence.

\section{Aesthetic-Evaluation Prompt Templates}
\label{app:prompts}

Figure~\ref{fig:offline-prompt} presents the prompt used to construct $\mathcal D_{\mathrm{off}}$. At request time, the five variables in braces are replaced with the corresponding expert mean scores for the song, and the song audio is attached to the request.

\begin{figure*}[p]
\centering
\begin{lrbox}{\offlinepromptbox}
\begin{minipage}{0.98\textwidth}
\vspace{3pt}
\begin{verbatim}
You are an expert music critic.

Listen to a song and write a professional, detailed critique grounded in
audible musical phenomena while taking the human experts' ratings as given.
Return the response in Simplified Chinese.

The five evaluation dimensions are:
1. Overall Coherence
   Evaluates musical and emotional continuity across the intro, verses,
   choruses, and outro. High ratings reflect smooth transitions, consistent
   dynamics, and a unified emotional tone throughout the piece.
2. Memorability
   Evaluates distinctive features--such as a catchy melody, rhythmic motif,
   or lyrical hook--that make the song easy to remember after one listen.
3. Naturalness of Vocal Breathing and Phrasing
   Evaluates phrasing quality and breath control, including alignment with
   semantic boundaries and rhythmic cues and whether breathing supports a
   fluent delivery without disrupting the singing flow.
4. Clarity of Song Structure
   Evaluates whether recognizable sections (e.g., verse, chorus, bridge) are
   clearly delineated and logically organized. Conventional and novel forms
   may both rate highly when their segmentation is musically meaningful.
5. Overall Musicality
   Evaluates listening enjoyment based on melody, harmony, instrumentation,
   and the integration of vocals and accompaniment.

Requirements:
1. Use professional terminology from music theory and audio production.
2. Analyze each dimension separately. Begin each analysis with its number and
   name, e.g., "1. Overall Coherence: ".
3. Ground each analysis in concrete musical evidence consistent with the given
   rating, such as melodic repetition/variation, section-level energy,
   rhythmic density, harmonic progression, vocal breaths, or arrangement.
4. Do not reassess the ratings. Explain them through professional, specific,
   and interpretable comments based on audible musical phenomena.
5. Do not state numerical ratings in the critique or say that a phenomenon
   occurs "because the rating is high/low."
6. Discuss only the song. Do not invent the singer's or band's identity.
7. The expert team's ratings (1--5) are:
   - Coherence: {coherence}
   - Memorability: {memorability}
   - Naturalness: {naturalness}
   - Clarity: {clarity}
   - Musicality: {musicality}
   After the critique, reproduce these ratings in a JSON code block. The scores
   are means across annotators and may therefore be non-integer values.
8. Ratings express agreement on a five-point Likert scale. Infer musical
   evidence according to the following mapping:
   - 5 (Excellent): outstanding quality with no discernible shortcomings.
   - 4 (Good): strong quality with only minor room for improvement.
   - 3 (Fair): adequate but unremarkable baseline quality.
   - 2 (Poor): clear shortcomings that impair the overall effect.
   - 1 (Very Poor): severe problems that fail to meet the criterion.

Output format:
Return valid JSON following this format, without an introduction or summary.
{
  "critical": "Evaluate the song along all five dimensions...",
  "scores": {
    "coherence": {coherence},
    "memorability": {memorability},
    "naturalness": {naturalness},
    "clarity": {clarity},
    "musicality": {musicality}
  }
}

Now evaluate the song.
\end{verbatim}
\vspace{2pt}
\end{minipage}
\end{lrbox}
\setlength{\fboxsep}{2pt}
\fcolorbox{gray!75!black}{gray!5!white}{\usebox{\offlinepromptbox}}
\caption{Prompt used by Gemini-3-Pro to generate offline aesthetic critiques. At request time, the five variables in braces are replaced with the corresponding expert mean scores, and the audio is supplied concurrently. The expert scores constrain the evaluative polarity of the critique for each dimension, while the prompt explicitly prohibits numerical ratings in the critique text.}
\label{fig:offline-prompt}
\end{figure*}

Figure~\ref{fig:model-prompt} presents the user prompt used for both the $\mathcal D_{\mathrm{on}}$ and $\mathcal D_{\mathrm{off}}$  training examples. This prompt is used for MuseCritic critique SFT, online critique generation, and subsequent reward modeling. The \texttt{<audio>} marker indicates the position at which the model processor inserts the song audio.

\begin{figure*}[p]
\centering
\begin{lrbox}{\onlinepromptbox}
\begin{minipage}{0.98\textwidth}
\vspace{3pt}
\begin{verbatim}
You are an expert music critic.

Listen to a song and write a professional, detailed critique in Simplified
Chinese, grounded in audible musical phenomena.

The five evaluation dimensions are:
1. Overall Coherence
   Evaluates musical and emotional continuity across the intro, verses,
   choruses, and outro. High ratings reflect smooth transitions, consistent
   dynamics, and a unified emotional tone throughout the piece.
2. Memorability
   Evaluates distinctive features--such as a catchy melody, rhythmic motif,
   or lyrical hook--that make the song easy to remember after one listen.
3. Naturalness of Vocal Breathing and Phrasing
   Evaluates phrasing quality and breath control, including alignment with
   semantic boundaries and rhythmic cues and whether breathing supports a
   fluent delivery without disrupting the singing flow.
4. Clarity of Song Structure
   Evaluates whether recognizable sections (e.g., verse, chorus, bridge) are
   clearly delineated and logically organized. Conventional and novel forms
   may both rate highly when their segmentation is musically meaningful.
5. Overall Musicality
   Evaluates listening enjoyment based on melody, harmony, instrumentation,
   and the integration of vocals and accompaniment.

Requirements:
1. Use professional terminology from music theory and audio production.
2. Analyze each dimension separately. Begin each analysis with its number and
   name, e.g., "1. Overall Coherence: ".
3. Ground each dimension in concrete musical evidence, such as melodic
   repetition/variation, section-level energy, rhythmic density, harmonic
   progression, vocal breaths, or changes in arrangement.
4. Provide professional, specific, and interpretable comments based on audible
   musical phenomena.
5. Discuss only the song. Do not invent the singer's or band's identity.

Now evaluate the song.
<audio>
\end{verbatim}
\vspace{2pt}
\end{minipage}
\end{lrbox}
\setlength{\fboxsep}{2pt}
\fcolorbox{gray!75!black}{gray!5!white}{\usebox{\onlinepromptbox}}
\caption{Prompt used to train the SFT-Critique-Generator, generate online aesthetic critiques with MOSS-Audio, and perform MuseCritic reward-model training. The model generates critiques along five dimensions without receiving expert scores; \texttt{<audio>} indicates the audio input position.}
\label{fig:model-prompt}
\end{figure*}

Figure~\ref{fig:judge-prompt} presents the shared scoring prompt used by Gemini-3.1-Pro for in-domain evaluation and by Qwen3-Omni for Music Arena evaluation. The song audio is inserted at the \texttt{<audio>} marker.

\begin{figure*}[p]
\centering
\begin{lrbox}{\judgepromptbox}
\begin{minipage}{0.98\textwidth}
\vspace{3pt}
\begin{verbatim}
You are an expert music critic.

Your task is to listen to a song, write a professional and detailed critique
grounded in audible musical phenomena, and assign final scores from 1 to 5
along five dimensions. Return the response in Simplified Chinese.

The five evaluation dimensions are defined as follows:
1. Overall Coherence
   This dimension evaluates the musical and emotional continuity across
   different sections of the song, including the intro, verse, chorus, and
   outro. High scores reflect smooth transitions, consistent dynamics, and a
   unified emotional tone throughout the piece.
2. Memorability
   This dimension evaluates distinctive musical features, such as a catchy
   melody, rhythmic motif, or lyrical hook, that make the song easy to remember
   after a single listen. Highly memorable songs typically exhibit a strong,
   repeatable musical identity.
3. Naturalness of Vocal Breathing and Phrasing
   This dimension evaluates phrasing quality and breath control in the vocal
   performance. It considers whether phrasing aligns with semantic boundaries
   and rhythmic cues and whether breathing supports a fluent, natural delivery
   without disrupting the singing flow.
4. Clarity of Song Structure
   This dimension evaluates how clearly the song is organized into recognizable
   sections, such as verse, chorus, and bridge, as well as the logic and
   coherence of the structural design. Both conventional and well-executed
   novel structures can receive high scores when segmentation is clear and
   musically meaningful.
5. Overall Musicality
   This dimension evaluates overall listening enjoyment based on melody,
   harmony, instrumentation, and the integration of vocals and accompaniment.
   It reflects the general aesthetic satisfaction derived from the song.

Task requirements:
For the provided song audio, follow these requirements:
1. Use professional terminology from music theory and audio production.
2. Analyze each dimension separately and in detail. Begin each analysis with
   its number and dimension name, for example, "1. Overall Coherence: ".
3. Ground every analysis in specific reasons and audible musical phenomena,
   such as melodic repetition or variation, section-level energy changes,
   changes in rhythmic density, harmonic progression, vocal breath placement,
   and changes in accompaniment or arrangement.
4. Provide professional, specific, and interpretable comments based on audible
   musical phenomena.
5. Discuss only the song itself. Do not invent the name of the singer or band.
   Only the song audio is provided.
6. At the end of the critique, provide the final scores. A higher score means
   that the song better satisfies the corresponding criterion. Use this format:
   Final Score:{"Coherence": score, "Musicality": score,
   "Memorability": score, "Clarity": score, "Naturalness": score}
7. Before assigning each score, use the following score-interval mapping and
   derive the final score strictly from the corresponding critique. Decimal
   scores are allowed:
   - (4,5] (Excellent): perfect performance on the dimension, with outstanding
     effectiveness and quality and no discernible flaws.
   - (3,4] (Good): strong performance on the dimension, with only minor room
     for improvement.
   - (2,3] (Fair): adequate but unremarkable baseline performance.
   - (1,2] (Poor): clear shortcomings on the dimension that impair the overall
     effect.
   - 1 (Very Poor): severe problems on the dimension that fail to satisfy its
     criterion.

Now evaluate the song.
<audio>
\end{verbatim}
\vspace{2pt}
\end{minipage}
\end{lrbox}
\setlength{\fboxsep}{2pt}
\fcolorbox{gray!75!black}{gray!5!white}{\usebox{\judgepromptbox}}
\caption{Shared prompt used for generative five-dimensional song scoring. Gemini-3.1-Pro uses this prompt for in-domain evaluation, and Qwen3-Omni uses it for Music Arena preference evaluation. Both models generate dimension-specific critiques and scores from 1 to 5; \texttt{<audio>} indicates the audio input position.}
\label{fig:judge-prompt}
\end{figure*}

\begin{table*}[p]
\centering
\small
\setlength{\tabcolsep}{3.6pt}
\renewcommand{\arraystretch}{0.88}
\begin{tabular}{llrrrrrrrrr}
\toprule
& & \multicolumn{4}{c}{Audiobox Aesthetics $\uparrow$} & \multicolumn{5}{c}{SongEval $\uparrow$} \\
\cmidrule(lr){3-6}\cmidrule(lr){7-11}
Window & Model & CE & CU & PC & PQ & CO & MU & ME & CL & NA \\
\midrule
\multirow{3}{*}{Full audio}
 & Muse & 6.79 & 7.50 & 5.77 & 7.93 & 3.76 & 3.56 & 3.72 & 3.65 & 3.62 \\
 & \musedirect & 6.87 & 7.50 & 5.90 & 7.93 & 3.85 & 3.64 & 3.82 & 3.73 & 3.68 \\
 & \musegrpo & \textbf{7.03} & \textbf{7.55} & \textbf{6.05} & \textbf{7.98} & \textbf{3.89} & \textbf{3.68} & \textbf{3.84} & \textbf{3.77} & \textbf{3.73} \\
\midrule
\multirow{3}{*}{360 s}
 & Muse & 6.93 & 7.53 & 5.92 & 7.97 & 3.85 & 3.65 & 3.81 & 3.73 & 3.70 \\
 & \musedirect & 6.98 & 7.52 & 6.01 & 7.96 & 3.91 & 3.70 & 3.87 & 3.78 & 3.73 \\
 & \musegrpo & \textbf{7.10} & \textbf{7.56} & \textbf{6.13} & \textbf{8.00} & \textbf{3.94} & \textbf{3.73} & \textbf{3.89} & \textbf{3.82} & \textbf{3.77} \\
\midrule
\multirow{3}{*}{330 s}
 & Muse & 7.00 & 7.55 & 6.00 & 7.99 & 3.88 & 3.69 & 3.83 & 3.76 & 3.72 \\
 & \musedirect & 7.05 & 7.54 & 6.09 & 7.98 & 3.94 & 3.73 & 3.90 & 3.81 & 3.75 \\
 & \musegrpo & \textbf{7.16} & \textbf{7.58} & \textbf{6.19} & \textbf{8.02} & \textbf{3.96} & \textbf{3.76} & \textbf{3.91} & \textbf{3.84} & \textbf{3.79} \\
\midrule
\multirow{3}{*}{300 s}
 & Muse & 7.10 & 7.58 & 6.11 & 8.03 & 3.92 & 3.73 & 3.87 & 3.80 & 3.76 \\
 & \musedirect & 7.15 & 7.57 & 6.20 & 8.02 & 3.97 & 3.78 & 3.93 & 3.85 & 3.79 \\
 & \musegrpo & \textbf{7.24} & \textbf{7.60} & \textbf{6.29} & \textbf{8.04} & \textbf{3.99} & \textbf{3.80} & \textbf{3.94} & \textbf{3.87} & \textbf{3.82} \\
\midrule
\multirow{3}{*}{270 s}
 & Muse & 7.21 & 7.60 & 6.24 & 8.06 & 3.96 & 3.78 & 3.90 & 3.84 & 3.79 \\
 & \musedirect & 7.27 & 7.61 & 6.32 & 8.06 & 4.01 & 3.82 & \textbf{3.97} & 3.89 & 3.83 \\
 & \musegrpo & \textbf{7.34} & \textbf{7.63} & \textbf{6.40} & \textbf{8.08} & \textbf{4.03} & \textbf{3.84} & \textbf{3.97} & \textbf{3.91} & \textbf{3.85} \\
\midrule
\multirow{3}{*}{240 s}
 & Muse & 7.33 & 7.64 & 6.37 & 8.11 & 4.01 & 3.84 & 3.94 & 3.89 & 3.83 \\
 & \musedirect & 7.38 & 7.65 & 6.44 & 8.10 & 4.05 & 3.87 & 3.99 & 3.92 & 3.86 \\
 & \musegrpo & \textbf{7.45} & \textbf{7.66} & \textbf{6.52} & \textbf{8.12} & \textbf{4.07} & \textbf{3.89} & \textbf{4.00} & \textbf{3.94} & \textbf{3.89} \\
\bottomrule
\end{tabular}
\caption{\textbf{Robustness of downstream aesthetic evaluation across audio windows.} All three policies are evaluated on the same 100 Muse test prompts. ``Full audio'' uses each complete generation; the remaining rows apply a common maximum-duration prefix, leaving shorter songs unchanged. \musedirect omits critiques during reward learning and scoring, whereas \musegrpo conditions its reward on self-generated critiques. Metric abbreviations follow Table~\ref{tab:downstream}; bold marks the best result within each window.}
\label{tab:grpo-window-robustness}
\end{table*}

\section{Music Arena Evaluation Details}
\label{app:arena-details}

Each Music Arena example contains a chosen and rejected song generated under the same text condition.
For \method and SongEval with the officially released weights \citep{yao2025songeval}, we average the five predicted aesthetic scores for each song.
We likewise average the four dimensions from Audiobox Aesthetics \citep{tjandra2025audiobox}, the five ratings generated by Qwen3-Omni \citep{xu2025qwen3omni}, and the seven dimensions predicted by the reward model trained by SongBench \citep{wu2026songbench}.
TuneJury \citep{kim2026tunejury} outputs one scalar preference score, which we use directly.
Under our evaluation configuration, CMI-RM \citep{ma2026cmirewardbench} provides a one-dimensional musicality score, which we use directly for pairwise comparison.
An example is correct only if $r(x^{+})>r(x^{-})$; ties count as errors.
Final accuracy is the fraction of correct predictions across all 733 preference pairs.

\section{Downstream GRPO Training and Evaluation Details}
\label{app:grpo-details}

\paragraph{Training data.}
We sample 500 multi-turn song-generation examples from the public Muse training data with seed 42 while preserving the original language distribution: 305 English and 195 Chinese examples.
All inputs retain the Muse multi-turn prompt format, and the policy is initialized from Muse-0.6B.

\paragraph{GRPO configuration.}
We train two policies from Muse-0.6B: \musedirect uses the w/o critique reward model, and \musegrpo uses the full \method reward model.
The two reward models share the same critique-SFT initialization, but the reward model used by \musedirect omits critiques during reward learning and scoring.
Training uses six NVIDIA H200 GPUs, a per-device batch size of 2, gradient accumulation over 2 steps, eight candidate generations, a learning rate of $10^{-6}$, a 5\% warmup, and one epoch.
The vLLM rollout server uses two NVIDIA H200 GPUs, a maximum model length of 20,000 and GPU memory utilization of 0.7.
Sampling uses temperature 0.9, top-$p$ 0.9, and a repetition penalty of 1.3.
Each completion contains at most 3,000 tokens, and a complete multi-turn trajectory contains at most 20,000 tokens.
For each policy, the reward for a candidate song is the mean of the five scores produced by its corresponding reward model; within-group relative advantages are computed according to GRPO.

\paragraph{Evaluation protocol.}
We evaluate on the 100 multi-turn test prompts from Muse \citep{jiang2026muse} and apply identical greedy decoding with $T=0$ to Muse, \musedirect, and \musegrpo.
Under modality-extended generation, deterministic decoding can exhibit inference-time instability on some evaluation examples: the model may fail to produce a valid ending and instead enter token-level repetition, producing an extended noisy tail.
This behavior does not consistently occur on a fixed subset of the evaluation prompts.
To prevent this decoding artifact from dominating the aesthetic assessment of otherwise valid song content while retaining all 100 test examples, we adopt a fixed-prefix protocol.
All generations longer than 240 seconds are evaluated using their first 240 seconds, whereas shorter generations remain unchanged and are neither padded nor looped.
The same deterministic preprocessing is applied to all three policies.
We then compute the nine metrics in Table~\ref{tab:downstream} with Audiobox Aesthetics and SongEval, focusing the main comparison on the aesthetic quality of the valid musical content.

\paragraph{Robustness across evaluation windows.}
As a robustness check, we evaluate both complete, untruncated generations and fixed-prefix windows of 240, 270, 300, 330, and 360 seconds.
Table~\ref{tab:grpo-window-robustness} shows that \musegrpo matches or exceeds both Muse and \musedirect on all nine metrics at every evaluation window at the reported precision, including the complete audio.
The consistent pattern across windows indicates that the downstream advantage of critique-conditioned rewards does not depend on the 240-second evaluation window used in the main results.

\section{Additional Experimental Results and Reproducibility}
\label{app:reproducibility}

\subsection{Inference-Time Interventions}
\label{app:intervention-details}

Table~\ref{tab:intervention-full} reports the complete results of the inference-time interventions described in Section~\ref{sec:ablation}.
For each mismatch run, the scoring stage receives the correct song audio but a critique randomly selected from another test song.
The no-audio condition retains the critique generated from the corresponding song but removes audio from the scoring stage.
All model parameters remain fixed.

\begin{table*}[p]
\centering
\scriptsize
\setlength{\tabcolsep}{4.2pt}
\renewcommand{\arraystretch}{0.72}
\begin{tabular}{llrrrr}
\toprule
Condition & Dimension & MSE $\downarrow$ & LCC $\uparrow$ & SRCC $\uparrow$ & KTAU $\uparrow$ \\
\midrule
\multirow{5}{*}{Matched}
 & Coherence & 0.2058 & 0.9147 & 0.8894 & 0.7262 \\
 & Memorability & 0.2511 & 0.9098 & 0.8839 & 0.7232 \\
 & Vocal naturalness & 0.2537 & 0.8906 & 0.8678 & 0.6945 \\
 & Structural clarity & 0.2275 & 0.9068 & 0.8806 & 0.7109 \\
 & Musicality & 0.2202 & 0.9120 & 0.8972 & 0.7344 \\
\midrule
\multirow{5}{*}{Mismatch 1}
 & Coherence & 0.3846 & 0.8253 & 0.8098 & 0.6208 \\
 & Memorability & 0.4858 & 0.8134 & 0.8019 & 0.6152 \\
 & Vocal naturalness & 0.4131 & 0.8089 & 0.7917 & 0.5984 \\
 & Structural clarity & 0.4061 & 0.8225 & 0.8080 & 0.6164 \\
 & Musicality & 0.4282 & 0.8217 & 0.8056 & 0.6135 \\
\midrule
\multirow{5}{*}{Mismatch 2}
 & Coherence & 0.3796 & 0.8278 & 0.8112 & 0.6251 \\
 & Memorability & 0.5171 & 0.8007 & 0.7841 & 0.6073 \\
 & Vocal naturalness & 0.4365 & 0.7974 & 0.7824 & 0.6013 \\
 & Structural clarity & 0.4282 & 0.8122 & 0.7967 & 0.6107 \\
 & Musicality & 0.4323 & 0.8206 & 0.8100 & 0.6297 \\
\midrule
\multirow{5}{*}{Mismatch 3}
 & Coherence & 0.3206 & 0.8568 & 0.8460 & 0.6662 \\
 & Memorability & 0.4239 & 0.8388 & 0.8211 & 0.6426 \\
 & Vocal naturalness & 0.3983 & 0.8158 & 0.8007 & 0.6139 \\
 & Structural clarity & 0.3874 & 0.8315 & 0.8216 & 0.6380 \\
 & Musicality & 0.3873 & 0.8406 & 0.8297 & 0.6453 \\
\midrule
\multirow{5}{*}{Mismatch 4}
 & Coherence & 0.3483 & 0.8434 & 0.8291 & 0.6526 \\
 & Memorability & 0.4792 & 0.8171 & 0.8087 & 0.6315 \\
 & Vocal naturalness & 0.4240 & 0.8046 & 0.7886 & 0.6023 \\
 & Structural clarity & 0.4313 & 0.8117 & 0.7969 & 0.6161 \\
 & Musicality & 0.4037 & 0.8336 & 0.8223 & 0.6396 \\
\midrule
\multirow{5}{*}{Mismatch 5}
 & Coherence & 0.4318 & 0.8026 & 0.7797 & 0.5948 \\
 & Memorability & 0.5501 & 0.7874 & 0.7626 & 0.5831 \\
 & Vocal naturalness & 0.4578 & 0.7877 & 0.7670 & 0.5824 \\
 & Structural clarity & 0.4618 & 0.7971 & 0.7776 & 0.5911 \\
 & Musicality & 0.4670 & 0.8045 & 0.7876 & 0.6037 \\
\midrule
\multirow{5}{*}{No audio}
 & Coherence & 0.9038 & 0.8641 & 0.8231 & 0.6369 \\
 & Memorability & 1.4011 & 0.8474 & 0.7966 & 0.6080 \\
 & Vocal naturalness & 0.8616 & 0.8327 & 0.7879 & 0.5985 \\
 & Structural clarity & 1.1359 & 0.8538 & 0.8007 & 0.6144 \\
 & Musicality & 0.8327 & 0.8512 & 0.8112 & 0.6301 \\
\bottomrule
\end{tabular}
\caption{\textbf{Complete results for inference-time critique and audio interventions.} ``Matched'' is the standard \method inference procedure. ``Mismatch 1--5'' denote five independent random permutations of critiques across the 200 test songs while retaining each song's audio. ``No audio'' retains the matched critique but removes audio during score prediction. Every intervention increases MSE and decreases LCC, SRCC, and KTAU relative to matched inference in all five dimensions.}
\label{tab:intervention-full}
\end{table*}

\subsection{Reproducibility Details}

We use random seed 42 for data splitting, model training, and sampling. 
Online critique generation, in-domain testing, Music Arena comparisons, and song generation for the 100 downstream prompts all use deterministic decoding to eliminate sampling variance in model comparisons. The fixed-prefix protocol described in Appendix~\ref{app:grpo-details} is used to handle occasional degenerate tails in downstream generations.

\paragraph{Hardware.}
Training and in-domain evaluation of \method and SongEval (UTMOS) use NVIDIA H200 GPUs. All model evaluations on Music Arena use NVIDIA H100 GPUs.

\section{Ethics Statement}
\label{app:ethics}

This study follows applicable data-privacy regulations and ethical standards in data acquisition and experimental evaluation.
Both the training and evaluation sets are constructed from public datasets, and the study uses no private personal data.
Given the data types and scope of this work, we identify no additional substantial ethical risks.

\paragraph{Potential risks.}
Like other learned aesthetic evaluators, \method may inherit genre, language, or stylistic preferences present in its training ratings and teacher-generated critiques.
We therefore recommend using \method as one component of a broader evaluation protocol and combining it with multiple evaluation metrics.

\section{AI Assistants in Research or Writing}
\label{app:ai-assistance}

In preparing this manuscript, AI assistants were employed solely to assist with refining the clarity, style, and readability of certain text segments.
They were not involved in designing the study, developing or implementing the methodology, collecting or analyzing data, or generating the primary scientific contributions.
All substantive research decisions, analyses, and conclusions are fully the responsibility of the authors.

\end{document}